\documentclass[aps,twocolumn,prl]{revtex4-1}

\usepackage{graphicx}
\usepackage{dcolumn}
\usepackage{bm}
\usepackage{xcolor}
\usepackage{amssymb}   
\usepackage{amsmath}
\usepackage{ulem}

\begin{document}

\title{
Emergent exotic superconductivity in artificially-engineered tricolor Kondo superlattices
}

\author{M. Naritsuka$^1$}
\author{T. Ishii$^1$}
\author{S. Miyake$^1$}
\author{Y. Tokiwa$^{1,2}$}
\author{R. Toda$^1$}
\author{M.~Shimozawa$^3$}
\author{T. Terashima$^1$}
\author{T. Shibauchi$^4$}
\author{Y. Matsuda$^1$}
\author{Y. Kasahara$^1$}

\affiliation{$^1$Department of Physics, Kyoto University, Kyoto 606-8502, Japan}
\affiliation{$^2$Experimentalphysik VI, Center for Electronic Correlations and Magnetism, Augsburg University, 86159 Augsburg, Germany}
\affiliation{$^3$Institute for Solid State Physics, University of Tokyo, Kashiwa 277-8581, Japan}
\affiliation{$^4$Department of Advanced Materials Science, University of Tokyo, Chiba 277-8561, Japan}

\begin{abstract}
In the quest for exotic superconducting pairing states, the Rashba effect, which lifts the electron-spin degeneracy as a consequence of strong spin-orbit interaction (SOI) under broken inversion symmetry, has attracted considerable interest. Here, to introduce the Rashba effect into two-dimensional (2D) strongly correlated electron systems,  we fabricate  non-centrosymmetric (tricolor) superlattices composed of three kinds of $f$-electron compounds with atomic thickness; $d$-wave heavy fermion superconductor CeCoIn$_5$ sandwiched by two different nonmagnetic metals, YbCoIn$_5$ and YbRhIn$_5$. We find that the Rashba SOI induced global inversion symmetry breaking in these tricolor Kondo superlattices  leads to profound changes in the superconducting properties of CeCoIn$_5$, which are revealed by unusual temperature and angular dependences of upper critical fields that are in marked contrast with the bulk CeCoIn$_5$ single crystals.  We demonstrate that the Rashba effect  incorporated into 2D CeCoIn$_5$ block layers is largely tunable by changing the  layer thickness.  Moreover, the temperature dependence of in-plane upper critical field exhibits an anomalous upturn at low temperatures, which is attributed to a possible emergence of a helical or stripe superconducting phase. Our results demonstrate  that the tricolor Kondo superlattices provide a new playground  for exploring exotic superconducting states in the strongly correlated 2D electron systems with the Rashba effect.
\end{abstract}

\maketitle

\section{Introduction}

Spin-orbit interaction (SOI) is a relativistic effect that entangles the spin and orbital degrees of freedom of the electrons.  Recently, it has been shown that the strong SOI in the presence of inversion symmetry breaking can dramatically affect the electronic properties, giving rise to a number of intriguing phenomena in various fields of contemporary condensed matter physics, such as spintronics \cite{Murakami,Sinova}, topological matter \cite{Qi}, and exotic superconductivity \cite{Bauer2012}.   In superconductors, the inversion symmetry imposes an important constraint on the pairing states.  In the presence of inversion symmetry, Cooper pairs are classified into spin-singlet and triplet states.  On the other hand, in the absence of inversion symmetry, an asymmetric potential gradient $\nabla V$ yields an SOI that gives rise to a parity-violated superconductivity \cite{Bauer2012}. Such a superconducting state exhibits unique properties including the admixture of spin-singlet and triplet states \cite{Gorkov,Frigeri}, unusual paramagnetic \cite{Frigeri} and electromagnetic response \cite{Mineev,Lu}, and topological superconducting states \cite{Qi,Tanaka,Sato}, which cannot be realized in conventional superconductors with global inversion symmetry. For instance, asymmetry of the potential in the direction perpendicular to the two-dimensional (2D) plane $\nabla V\parallel [001]$ induces the Rashba SOI $\alpha_\mathrm{R}\bm{g}(\bm{k})\cdot\bm{\sigma}\propto(\bm{k}\times\nabla V)\cdot\bm{\sigma}$, where $\bm{g}(\bm{k})=(k_y,-k_x,0)/k_\mathrm{F}$, $k_\mathrm{F}$ is the Fermi wave number, and $\bm{\sigma}$ is the Pauli matrix. The Rashba SOI splits the Fermi surface into two sheets with different spin structures \cite{Bauer2012,Rashba,Bychkov}.  The energy splitting is given by $\alpha_\mathrm{R}$, and the spin direction is tilted into the plane, rotating clockwise on one sheet and anticlockwise on the other. When the Rashba splitting  exceeds the superconducting gap energy ($\alpha_\mathrm{R}>\Delta$), the superconducting state is dramatically modified. 

It has been pointed out theoretically that the effect of the Rashba SOI on superconductivity can be more pronounced 
by strong electron correlations \cite{Fujimoto,Maruyama2015}.  Although strongly correlated heavy fermion superconductors with broken inversion symmetry, such as CePt$_3$Si \cite{Bauer}, CeRhSi$_3$ \cite{Kimura}, and UIr \cite{Akazawa}, have been reported,  the superconductivity often coexists with magnetic order in these compounds. Moreover, the magnitude of the Rashba SOI is hard to control, as it is determined by the crystal structure.  In heavy transition metal oxides with 4$d$ or 5$d$ elements, it has been suggested that the cooperative effect of the strong electron correlation and strong SOI gives rise to exotic electronic states such as Weyl semimetal and topological Mott insulator \cite{Krempa}, but superconductivity in such materials has not been reported so far.

 \begin{figure*}[t]
	\begin{center}
		\includegraphics[width=0.7\linewidth]{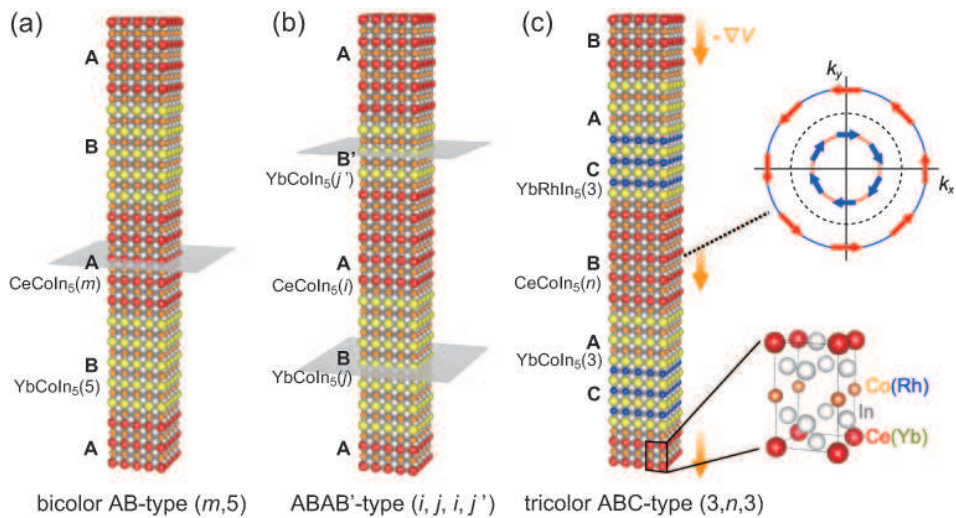}
		\caption{
		Schematic representations of Kondo superlattices. (a) Centrosymmetric bicolor superlattices CeCoIn$_5$($m$)/YbCoIn$_5$(5). The center of a CeCoIn$_5$ BL (ash plane) is a mirror plane. (b) $ABAB^\prime$-type superlattices CeCoIn$_5$($i$)/YbCoIn$_5$($j$)/CeCoIn$_5$($i$)/YbCoIn$_5$($j^\prime$) ($j\neq j^\prime$).  The center of a CeCoIn$_5$ BL is not a mirror plane, but the centers of YbCoIn$_5$ BLs (ash planes) are mirror planes. 
		(c) Non-centrosymmetric tricolor superlattices YbCoIn$_5$(3)/CeCoIn$_5$($n$)/YbRhIn$_5$(3). 
		Crystal structure of $TM$In$_5$ ($T$: Ce or Yb, $M$: Co or Rh) with tetragonal symmetry is also shown. In the tricolor superlattices, all layers are not the mirror planes. The orange arrows represent the asymmetric potential gradient $-\nabla V$ due to the broken  inversion symmetry.  The Rashba SOI splits the Fermi surface into two sheets; spin direction rotates clockwise on one sheet (blue arrows) and anticlockwise on the other (red arrows).  
		}
	\end{center}
\end{figure*}

Recently, the Rashba effect is a topic of growing interest in 2D superconductivity on the surface of the substrate and at the interface between  two different materials, which necessarily have broken inversion symmetry.  In some of these systems the Rashba-type SOI is tunable and enables the possibility of achieving exotic states such as  topological superconducting states.  However, in these 2D systems discovered until now,  superconductivity emerges from weakly correlated electron states \cite{Reyren,Ueno,Qin,Wang,Saito,Nam}.  Thus in the superconductors with strong Rashba SOI, the role of strong electron-electron interaction has remained largely unexplored due to the lack of suitable material systems.  The situation calls for new 2D systems with significant electron correlations and tunable Rashba SOI, which make such investigations possible.

Recent technological advances in fabricating Kondo superlattices, where a Ce-based heavy-fermion compound and a nonmagnetic conventional metal are stacked alternatively, open up new playgrounds for investigating 2D strongly correlated superconductors \cite{Shishido2010,Shimozawa2016}.  The SOI in Ce-based compounds is generally significant because of heavy elements. It has been demonstrated that in  CeCoIn$_5$/YbCoIn$_5$ superlattices illustrated in Fig.\,1(a), where a strongly correlated heavy-fermion superconductor CeCoIn$_5$ \cite{Petrovic} and nonmagnetic metal YbCoIn$_5$ \cite{Huy} are stacked alternately as ``$ABAB\cdots$" the superconducting heavy quasiparticles as well as magnetic fluctuations can be confined within Ce block layers (BLs) with atomic thickness \cite{Mizukami,Goh,Yamanaka}. These ``bicolor" superlattices maintain centrosymmetry, although it has been suggested that the local inversion symmetry breaking at the interface between two compounds influences the superconducting and magnetic properties \cite{Goh,Yamanaka,Ishii,Maruyama2012}. It has been shown that the effect of inversion symmetry breaking appears to be pronounced in $ABAB^\prime$-type superlattices, where $i$-unit-cell-thick (UCT) CeCoIn$_5$ is sandwiched by $j$- and $j^\prime$-UCT YbCoIn$_5$ ($j\neq j^\prime$) [Fig.\,1(b)] \cite{Shimozawa2014}. In these superlattices, inversion symmetry is not preserved in CeCoIn$_5$ BLs owing to the thickness modulation of the YbCoIn$_5$ BLs, in addition to the local inversion symmetry breaking. However, as shown in Fig.\,1(b), mirror planes are present in the YbCoIn$_5$ BLs and hence the global inversion symmetry is preserved in the whole crystals. Therefore, it is still a challenging issue to realize exotic superconducting phenomena associated with the global inversion symmetry breaking in the Kondo superlattices.

Here, to introduce a global inversion symmetry breaking in these Kondo superlattices, we  fabricated ``tricolor" Kondo superlattice with an asymmetric sequence YbCoIn$_5$/CeCoIn$_5$/YbRhIn$_5$, in which CeCoIn$_5$ is sandwiched by two different nonmagnetic metals, YbCoIn$_5$ and YbRhIn$_5$, as illustrated in Fig.~1(b). These tricolor superlattices with an asymmetric ``$ABCABC\cdots$" arrangement introduces broken inversion symmetry along the stacking direction.  Although the crystal structure of bulk CeCoIn$_5$ possesses the inversion symmetry, band structure calculations suggest that even a small degree of inversion symmetry breaking can induce a large Rashba splitting of the Fermi surface \cite{Shimozawa2014}.  We demonstrate that the Rashba SOI  induced in the tricolor Kondo superlattices  leads to profound changes in the nature of the superconductivity in CeCoIn$_5$.  Moreover, by changing the thickness of CeCoIn$_5$ BL, the magnitude of the Rashba SOI is largely controllable.  We also show a possible emergence of an exotic superconducting phase in magnetic field applied parallel to the layers.

\begin{figure*}[t]
	\begin{center}
		\includegraphics[width=0.9\linewidth]{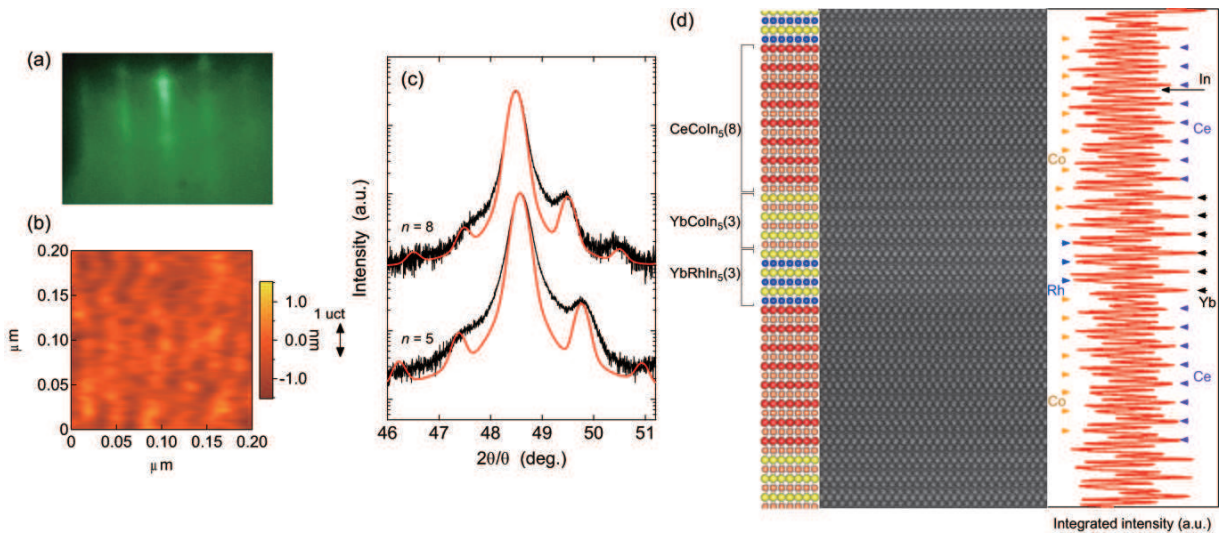}
		\caption{ 
		(a) Typical RHEED streak patterns for $n=8$ tricolor superlattice taken during the crystal growth. 
		(b) Typical AFM image for $n=8$ tricolor superlattice grown by MBE. 
		(c) Cu K$\alpha_1$ X-ray diffraction patterns for $n = 5$ and 8 	superlattices around the main (004) peaks. Red lines represent the step-model simulations ignoring the interface and layer-thickness fluctuations.
		(d) High-resolution cross-sectional TEM image of the $n=8$ tricolor superlattice with the electron beam aligned along the (110) direction. The right panel is the intensity integrated over the horizontal width of the image. The positions of Ce, Yb, Co, and Rh atoms can be identified from the dips and peaks as shown by arrows, which are consistent with the designed superlattice structure shown in the left panel. 
		}
	\end{center}
	\vspace{-5mm}
\end{figure*}

\section{Experimental Method}
The tricolor Kondo superlattices (YbCoIn$_5$/CeCoIn$_5$/YbRhIn$_5$)$_\ell$ with $c$-axis oriented structure are epitaxially grown on MgF$_2$ substrate using the molecular beam epitaxy (MBE) technique.  We first grow CeIn$_3$ ($\sim20$\,nm) as a buffer layer on MgF$_2$.  Then  3 UCT YbCoIn$_5$, $n$-UCT CeCoIn$_5$ ($n=5$ and 8), and 3-UCT YbRhIn$_5$ are grown alternatively. The sequence of YbCoIn$_5$(3)/CeCoIn$_5$($n$)/YbRhIn$_5$(3) is stacked repeatedly $\ell=40$ times for $n=5$ and $\ell=30$ times for $n=8$ tricolor superlattices,  so that the total thickness is about 300\,nm. We also fabricate bicolor superlattices consisting of alternating layers of  $m$-UCT CeCoIn$_5$ ($m=5$ and 8) and 5-UCT YbCoIn$_5$, CeCoIn$_5$($m$)/YbCoIn$_5$(5), and those of $5$-UCT CeCoIn$_5$ and 5-UCT YbRhIn$_5$, CeCoIn$_5$(5)/YbRhIn$_5$(5).

The crystalline quality of  tricolor superlattices was evaluated by several techniques.   Streak pattern of  the reflection high-energy electron diffraction (RHEED) image shown in Fig.\,2(a) was observed during the whole growth of the superlattices, indicating good epitaxy.  The atomic force microscope (AFM) image shown in Fig.\,2(b) reveals that the surface roughness is within $\pm1$\,nm, which is comparable to one UCT along the $c$ axis of the constituents (CeCoIn$_5$, YbCoIn$_5$, and YbRhIn$_5$). The atomically flat regions extend over distances of $\sim0.1$\,$\mu$m, showing that the transport properties are not expected to be seriously influenced by the roughness. The X-ray diffraction patterns are shown in Fig.\,2(c).  The position of the lateral satellite peaks and their asymmetric heights can be reproduced by the step-model simulations (red lines) neglecting interface and layer-thickness fluctuations.  This indicates the growth with no discernible interdiffusion across the interfaces, confirming that the superlattices were fabricated as designed. Figure\,2(d) depicts the high-resolution cross-sectional transmission electron microscope (TEM) image along the (110) direction for $n=8$ tricolor superlattice.  Clear interface between CeCoIn$_5$ and YbRhIn$_5$/YbCoIn$_5$ layers is observed.  The intensity integrated along the horizontal direction of the image plotted against vertical position shows that the intensities of Ce, Yb, Co, Rh, and In atoms are almost constant within each BLs. A clear difference between the Ce and Yb layers can be seen in the integrated intensity. Although the interface between YbCoIn$_5$ and YbRhIn$_5$ is less visible in the TEM image, there is a discernible difference in the integrated intensity at Co and Rh sites in YbRhIn$_5$/YbCoIn$_5$ layers. The sharp boundaries at the interfaces between BLs also demonstrate the absence of the interdiffusion. Because we have 3-UCT YbCoIn$_5$ and 3-UCT YbRhIn$_5$ spacers between CeCoIn$_5$, the Ruderman-Kittel-Kasuya-Yoshida interaction between the adjacent Ce BLs is less than 0.1\,\% of that between the neighboring Ce atoms in the same layer \cite{Peters}, indicating that CeCoIn$_5$ BLs are magnetically decoupled. Moreover, the superconducting proximity effect between CeCoIn$_5$ layer and neighboring YbCo(Rh)In$_5$ layer is expected to be negligibly small due to the large Fermi velocity mismatch \cite{She}. In fact, it has been shown that in the superlattices with 4-6 UCT CeCoIn$_5$ layers, whose thickness is comparable to the perpendicular coherence length $\xi_c\sim3$-4\,nm, 2D heavy fermion superconductivity is realized \cite{Mizukami,Goh,Schneider}.

\section{Results and Discussion}
\subsection{Rashba spin-orbit interaction in tricolor Kondo superlattices}

 Figure~3 depicts the temperature ($T$) dependence of the resistivity $\rho(T)$ for $n=5$ and 8 tricolor superlattices.  Superconducting transition is observed at $T_c=0.4$\,K and 0.8\,K (determined by the mid point of the resistive transition) for $n=5$ and 8, respectively.   For comparison, $\rho(T)$ for $m=5$ and 8 bicolor CeCoIn$_5$($m$)/YbCoIn$_5$(5) superlattices are also shown.  The transition temperatures of the tricolor superlattices are suppressed compared with those of the CeCoIn$_5$ thin film ($T_c=2.0$\,K) and the bicolor superlattices with the same CeCoIn$_5$ BL thickness.    This reduction of $T_c$ is unlikely due to the difference in the impurity and interface roughness scatterings, since the residual resistivity at $T\rightarrow0$  and residual resistivity ratio at $T_c$ [$\rho(300\,{\rm K})/\rho(T_c)$] of the bicolor and tricolor superlattices  are comparable. Moreover, $T_c$ of bicolor CeCoIn$_5$(5)/YbCoIn$_5$(5) and CeCoIn$_5$(5)/YbRhIn$_5$(5) superlattices is higher than $T_c$ of tricolor superlattice with the same CeCoIn$_5$ BL thickness (inset of Fig.\,3), implying that the strain effect at the interfaces is not important for determining $T_c$. We point out that the reduction of $T_c$ in the tricolor superlattices can be attributed to the Rashba effect.  In fact, the Fermi surface splitting due to the Rashba effect should modify seriously the nesting condition and hence is expected to reduce the commensurate antiferromagnetic (AFM) fluctuations with a wave vector ${\bm Q}=(\pi/a,\pi/a,/\pi/c)$, which is dominant in bulk CeCoIn$_5$ \cite{Stock}. In addition, the broken inversion symmetry reduces the AFM fluctuations by lifting the degeneracy of the fluctuation modes through the helical anisotropy of the spin configuration \cite{Yamanaka,Ishii,Yanase2008,Takimoto}.  Given that the superconductivity of CeCoIn$_5$ is mediated by AFM fluctuations,  the reduction of the AFM fluctuations lead to the suppression of $T_c$ in the tricolor superlattices.

\begin{figure}[t]
	\begin{center}
		\includegraphics[width=0.7\linewidth]{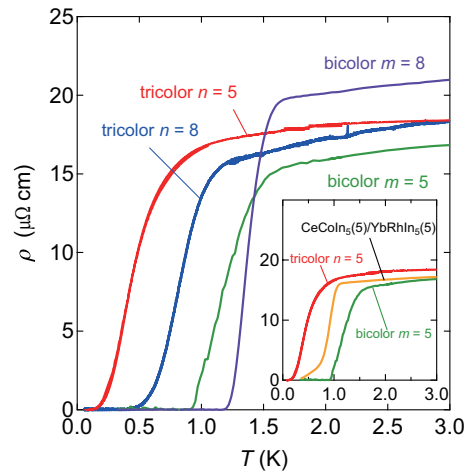}
		\caption{
		Temperature dependence of the resistivity $\rho(T)$ in the $n=5$ and 8 tricolor YbCoIn$_5$(3)/CeCoIn$_5$($n$)/YbRhIn$_5$(3) superlattices  and in the $m=5$ and 8 bicolor CeCoIn$_5$($m$)/YbCoIn$_5$(5) superlattices. Inset: $\rho(T)$ for the tricolor YbCoIn$_5$(3)/CeCoIn$_5$($n=5$)/YbRhIn$_5$(3), bicolor CeCoIn$_5$($m=5$)/YbCoIn$_5$(5), and bicolor CeCoIn$_5$(5)/YbRhIn$_5$(5) superlattices.
		}
	\end{center}
\end{figure}

Figures\,4(a) and 4(b) show the resistive transitions in magnetic fields applied parallel to the 2D plane for $n=8$ and 5  tricolor superlattice, respectively.   Figures\,4(c) and 4(d)  show the expanded view at low temperatures.     As shown in Figs.\,4(c) and 4(d), a nearly parallel shift of the resistive transition to lower temperatures with increasing magnetic field is observed in both superlattices. Figures\,4(e) and 4(f) depict in-plane $H_{c2}$ determined by four different criteria [$\rho(T,H)=0.3\rho_N(T)$, $0.5\rho_N(T)$, $0.7\rho_N(T)$, and $0.9\rho_N(T)$], as a function of temperature for $n=8$ and 5 tricolor superlattices, respectively. Here we discuss the anisotropy of upper critical field $H_{c2}$ of the tricolor superlattices.   Because of rather broad transition temperature width especially for $n=5$ tricolor superlattice, there is  ambiguity in determining $H_{c2}$.  We then defined  $H_{c2}(T)$  as the magnetic field at which the resistivity drops to 50\,\% of its normal state value $\rho(T, H)=0.5\rho_N(T)$.   Figure\,5(a) shows the anisotropy of upper critical field $H_{c2\parallel}/H_{c2\perp}$ plotted as a function of $T/T_c$, where $H_{c2\parallel}$ and $H_{c2\perp}$ are upper critical fields in magnetic fields parallel and perpendicular to the layer, respectively.   In sharp contrast to the CeCoIn$_5$ single crystal \cite{Tayama} and thin film with thickness of 120\,nm, the temperature dependence of $H_{c2\parallel}/H_{c2\perp}$ of the tricolor superlattices exhibits a diverging behavior upon approaching to $T_c$.  This diverging behavior is a characteristic feature of the 2D superconductivity, which is consistent with the thickness of CeCoIn$_5$ BL comparable or less than $\xi_c$.  We note that the diverging $H_{c2\parallel}/H_{c2\perp}$ near $T_c$ is observed even when $H_{c2}$ is defined by using $\rho(T,H)=0.3\rho_N(T)$, $0.7\rho_N(T)$, and $0.9\rho_N(T)$, indicating that the 2D nature of the superconductivity is irrespective of the determination of $H_{c2}$. The magnitude of $H_{c2\parallel}/H_{c2\perp}$ near $T_c$ for $n=5$ is smaller than that for $n = 8$, which is opposite to the expected behavior in conventional 2D superconductors. We point out that this reduction is explained by the suppression of the Pauli pair-breaking effect owing to the Rashba SOI, which will be discussed later. 

\begin{figure*}[t]
	\begin{center}
		\includegraphics[width=0.85\linewidth]{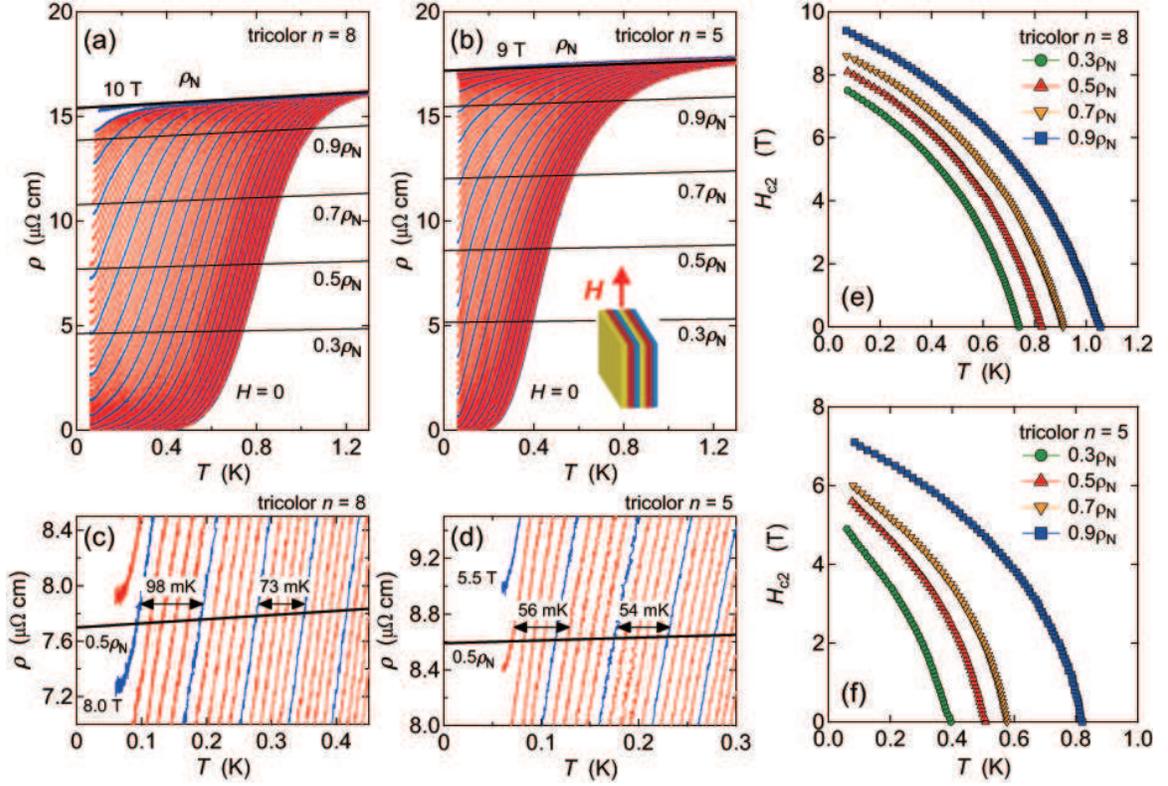}
		\caption{
		Temperature dependence of resistivity for (a) $n=8$ 	and (b) $n=5$ tricolor superlattices, respectively, in applied field parallel to the plane. Red and blue curves are taken in every 0.1 and 0.5~T, respectively. 
		Thin solid black curve represents the normal state resistivity $\rho_N$, and thin solid lines represent $0.3\rho_N$, $0.5\rho_N$, $0.7\rho_N$, and $0.9\rho_N$. 
		(c) and (d) are expanded views of (a) and (b) around $0.5\rho_N$, respectively. The arrows correspond to the shift of $T_c$ by changing the magnetic field of 0.5\,T. 
		(e) and (f) show in-plane upper critical field defined by using four different criteria, $\rho(T,H)=0.3\rho_N(T)$, $0.5\rho_N(T)$, $0.7\rho_N(T)$, and $0.9\rho_N(T)$, as a function of temperature for $n=8$ and 5 tricolor superlattices, respectively. 
		}
	\end{center}
\end{figure*}

\begin{figure*}[t]
	\begin{center}
		\includegraphics[width=0.7\linewidth]{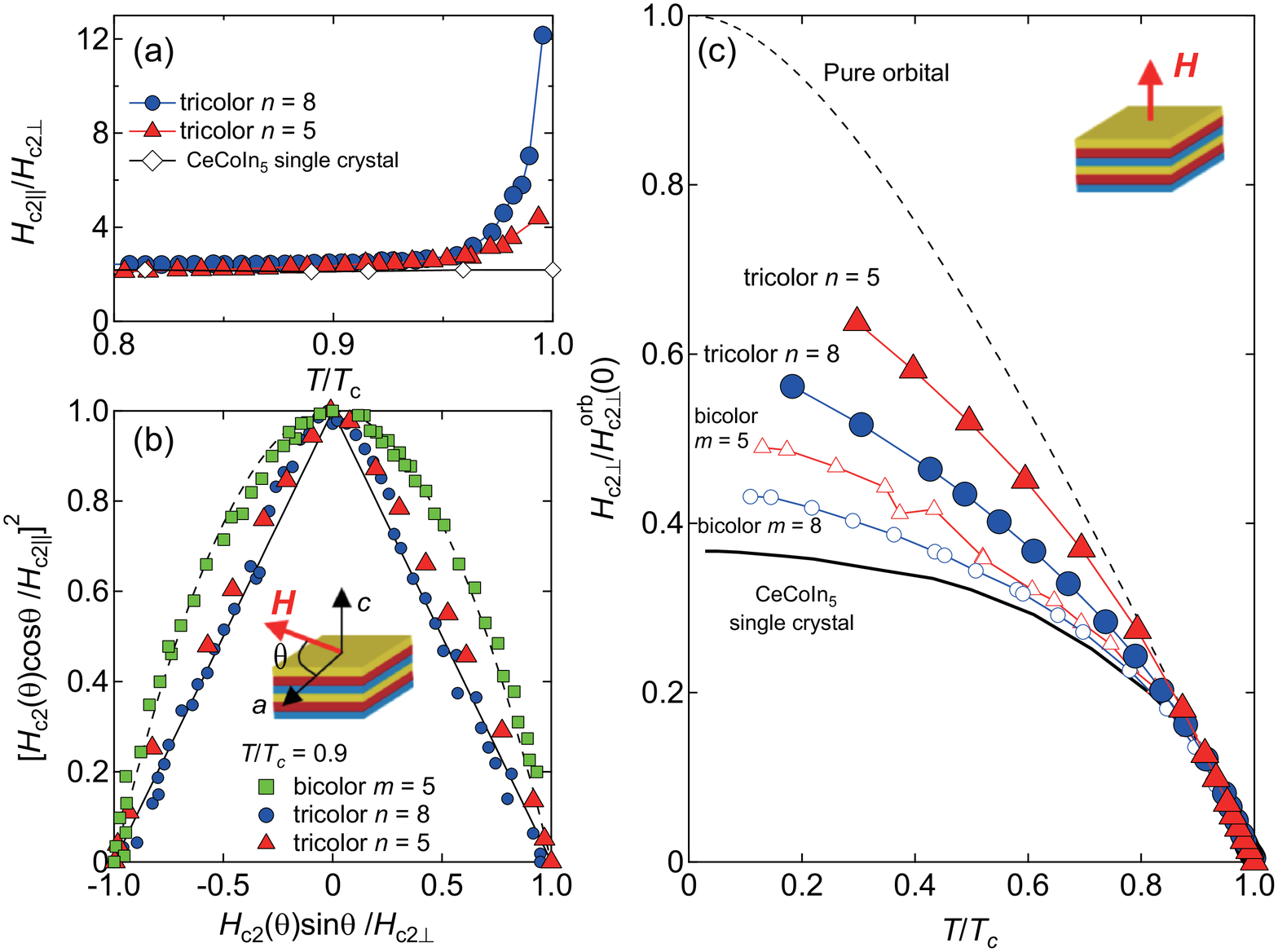}
		\caption{
		(a) The anisotropy of upper critical field $H_{c2}$, $H_{c2\parallel}/H_{c2\perp}$, plotted as a function of normalized temperature $T/T_c$ for the tricolor superlattices and CeCoIn$_5$ single crystal. 
		(b) Angular dependence of $H_{c2}$, $H_{c2}(\theta)$, plotted in an appropriate dimensionless form for the tricolor and bicolor CeCoIn$_5$($m$)/YbCoIn$_5$(5) superlattices. The solid and dashed lines represent the Tinkham's formula for a two-dimensional superconductor and the three-dimensional anisotropic mass model, which are described as $[H_{c2}(\theta)\cos\theta/H_{c2\parallel}]^2=-|H_{c2}(\theta)\sin\theta/H_{c2\perp}|+1$ \cite{Tinkham1963} and $[H_{c2}(\theta)\cos\theta/H_{c2\parallel}]^2=-[H_{c2}(\theta)\sin\theta/H_{c2\perp}]^2+1$ \cite{Tinkham}, respectively. 
		(c) Normalized upper critical field in perpendicular fields, $H_{c2\perp}/H_{c2\perp}^\mathrm{orb}(0)$, as a function of $T/T_c$ for the tricolor superlattices, compared with that for the bicolor CeCoIn$_5$($m$)/YbCoIn$_5$(5) superlattices with the same CeCoIn$_5$ block layer thickness. We also plot $H_{c2\perp}/H_{c2\perp}^\mathrm{orb}$ for CeCoIn$_5$ single crystal \cite{Tayama} with a strong Pauli pair-breaking effect and the WHH curve \cite{WHH} without the Pauli pair-breaking effect. In (a)-(c), $H_{c2}$ is defined by using $\rho(T,H)=0.5\rho_N(T)$.
		}
	\end{center}
\end{figure*}

The salient feature of the superconductivity in the tricolor superlattices, which is distinctly different from the bicolor superlattices, is revealed by the angular and temperature dependences of $H_{c2}$. Figure\,5(b) shows the angular dependence of the upper critical field $H_{c2}(\theta)$ for $n=5$ and 8 tricolor superlattices, where $\theta$ is the angle between {\boldmath $H$} and $ab$ plane.  For comparison,  $H_{c2}(\theta)$ of $m=5$ bicolor CeCoIn$_5$($m$)/YbCoIn$_5$(5) superlattice is also shown.   In Fig.\,5(b), square of the in-plane component of $H_{c2}$,  $[H_{c2}(\theta)\cos\theta]^2$, is plotted as a function of out-of-plane component, $H_{c2}(\theta)\sin\theta$.   A cusp appears in the angular dependence near parallel field in the tricolor superlattices, which is in sharp contrast to the result of $m=5$ bicolor superlattice that shows  smooth angular dependence with no cusp.   In 2D superconductor and  layered superconductors, where superconducting layer thickness $d$ is smaller than  $\xi_c$, $H_{c2}(\theta)$ exhibits a characteristic angle dependence.  When the superconductivity is destroyed by the orbital motion of Cooper pairs in the 2D plane, $H_{c2}(\theta)$ obeys the following equation derived by Tinkham \cite{Tinkham1963}:
\begin{equation}
[H_{c2}(\theta)\cos\theta/H_{c2\parallel}]^2=-|H_{c2}(\theta)\sin\theta/H_{c2\perp}|+1.
\end{equation}
Therefore $H_{c2}(\theta)$ is not differentiable at $\theta=0$  and follows a cusp-like dependence at small $\theta$.  On the other hand, when the superconductivity is dominated by Pauli paramagnetic pair-breaking effect,  the upper critical field is given by $H_\mathrm{P}=\sqrt{2}\Delta/g\mu_\mathrm{B}$ \cite{Clogston}, where $g$ is the gyromagnetic ratio, $\Delta$ is the superconducting gap amplitude, and $\mu_\mathrm{B}$ is the Bohr magneton.  Then angular dependence of $H_{c2}$ is determined by the anisotropy of $g$-factor, which is smooth for all $\theta$, and can be described by \cite{Tinkham}
\begin{equation}
[H_{c2}(\theta)\cos\theta/H_{c2\parallel}]^2=-[H_{c2}(\theta)\sin\theta/H_{c2\perp}]^2+1,
\end{equation}
similar to anisotropic mass model of anisotropic  3D superconductors. As shown by the solid and dashed lines in Fig.\,5(b), $H_{c2}(\theta)$ of $n=5$ and 8 tricolor superlattices are well fitted by Eq.(1), while $H_{c2}(\theta)$ of $m$\,=\,5 bicolor superlattice is well fitted by Eq.(2).  We note that  cusp-like behavior of $H_{c2}(\theta)$ near $\theta=0$ appears even when $H_{c2}$ is determined by using $\rho(T, H)=0.3\rho_N(T)$, $0.7\rho_N(T)$, and $0.9\rho_N(T)$, indicating the intrinsic properties of the tricolor superlattices.  These results strongly suggest that the Pauli paramagnetic pair-breaking effect is dominant in the bicolor superlattices, whereas the orbital pair-breaking effect is dominant in the tricolor superlattices.

 More direct evidence for the suppression of the Pauli pair-breaking effect in the tricolor superlattices is provided by  $H_{c2\perp}$ normalized by the orbital limited upper critical field without Pauli effect  $H_{c2\perp}^{\rm orb}$.  Figure\,5(c) displays $H_{c2\perp}(T)/H_{c2\perp}^{\rm orb}(0)$ of the bicolor and tricolor superlattices plotted 
as a function of $T/T_c$.  Here  $H_{c2\perp}^\mathrm{orb}(0)$ is calculated by the initial slope of $H_{c2\perp}$ at $T_c$ by using Werthamer-Helfand-Hohenberg (WHH) formula \cite{WHH}, $H_{c2\perp}^\mathrm{orb}(0)=-0.69T_c(dH_{c2\perp}/dT)_{T_c}$.  For comparison,  we include the two extreme cases, $H_{c2\perp}/H_{c2\perp}^\mathrm{orb}(0)$  for CeCoIn$_5$ single crystal \cite{Tayama}, in which superconductivity is dominated by Pauli pair-breaking effect \cite{Izawa,Tayama}, and the WHH curve without the Pauli effect.   For the bicolor superlattices, $H_{c2\perp}/H_{c2\perp}^\mathrm{orb}(0)$ is enhanced from that of bulk CeCoIn$_5$ only slightly, indicating that the superconductivity is still dominated by Pauli effect, consistent with the angular variation of $H_{c2}(\theta)$.   What is remarkable is that $H_{c2\perp}/H_{c2\perp}^\mathrm{orb}(0)$ of the tricolor superlattices is dramatically enhanced from that of the bicolor superlattices with the same CeCoIn$_5$ BL thickness.  In particular, $H_{c2\perp}/H_{c2\perp}^\mathrm{orb}(0)$ of $n=5$ tricolor superlattice is close to the WHH curve with no Pauli effect.  This is again consistent with $H_{c2}(\theta)$. The enhancement of $H_{c2\perp}/H_{c2\perp}^\mathrm{orb}(0)$ is attributed to the enhancement of the Pauli limiting field, leading to the increase of the relative importance of the orbital pair-breaking effect compared to the Pauli pair-breaking effect. 

There are several possible origins for the enhancement of the Pauli limiting field in the tricolor superlattices, including (1) reduction of $g$-value, (2) enhancement of $\Delta/T_c$, and  (3) the Rashba effect. We point out that both (1) and (2) are unlikely because of the following reasons.  Since the  $g$-value  is determined by the crystalline electric field,   the $g$-value of the tricolor superlattices should be close to the value of the bicolor superlattices with very similar crystal structure.  Moreover, in CeRhIn$_5$ with similar crystal and electronic structure, $g$-value is insensitive to the applied pressure \cite{Knebel}.   Recent site-selective-nuclear magnetic resonance experiments of the bicolor superlattices reveal that the AFM fluctuations in CeCoIn$_5$ BLs are suppressed with decreasing $m$ \cite{Yamanaka},  implying that  pairing interaction is expected to be weakened with decreasing CeCoIn$_5$ layer thickness.   However, this tendency is  opposite to the observed enhancement of Pauli limiting field in the bicolor and tricolor superlattices with thinner BL thickness. We stress that the dramatic suppression of the Pauli effect in the tricolor superlattices is naturally explained by the Rashba effect. In the presence of external magnetic field satisfying $\alpha_\mathrm{R}\gg\mu_\mathrm{B}H$, the quasiparticle energy dispersion in the Rashba system is given as \cite{Bauer2012,Samokhin}
 \begin{equation} E_\pm(\bm{k},\bm{H})\approx\xi(\bm{k})\pm\alpha_\mathrm{R}|\bm{k}|\mp\mu_\mathrm{B}\bm{g}(\bm{k})\cdot\bm{H},
 \end{equation}
 where $\xi(\bm{k})$ is the quasiparticle energy without Rashba term and magnetic field and $\alpha_\mathrm{R}|\bm{k}|$ is the Rashba type spin-orbit splitting. The Zeeman interaction given by $\mu_\mathrm{B}\bm{g}(\bm{k})\cdot\bm{H}$ leads to anisotropic suppression of the Pauli pair-breaking effect; Strong suppression of the Pauli effect occurs for $\bm{H}\parallel [001]$ where $\bm{g}(\bm{k})\cdot\bm{H}=0$ \cite{Gorkov,Frigeri,Mineev}, while the suppression is weaker for $\bm{H}\parallel ab$ since $\bm{g}(\bm{k})\perp\bm{H}$ is not always satisfied. Therefore, $H_{c2\perp}$ is more strongly enhanced than $H_{c2\parallel}$ by the Rashba effect, giving rise to the reduction of $H_{c2\parallel}/H_{c2\perp}$. Since the fraction of the noncentrosymmetric interface increases rapidly with decreasing $n$, the magnitude of $H_{c2\parallel}/H_{c2\perp}$ near $T_c$ for $n=5$ tricolor superlattices is smaller than that for $n=8$.

 Thus both the angular and temperature dependences of $H_{c2}$ indicate that the Pauli paramagnetic pair breaking effect, which is dominant not only in CeCoIn$_5$ single crystal but also  in the centrosymmetric bicolor superlattices,  can be substantially reduced in the tricolor superlattices.   Here we comment on  the slight enhancement of  $H_{c2\perp}/H_{c2\perp}^\mathrm{orb}(0)$ from bulk CeCoIn$_5$  value in the bicolor superlattices.  This  has been attributed to the local inversion symmetry breaking at the top and bottom interfaces of CeCoIn$_5$ layers at the immediate proximity to YbCoIn$_5$ BLs \cite{Goh,Maruyama2012}. In tricolor superlattices, $H_{c2\perp}/H_{c2\perp}^\mathrm{orb}(0)$ approaches orbital limit with decreasing $n$. This demonstrates that the Rashba SOI effect incorporated into 2D CeCoIn$_5$ BLs due to the built-in broken inversion symmetry is largely tunable by changing the BL thickness in the present tricolor superlattices.

 \subsection{A possible exotic superconducting state in parallel field}

Here we investigate the superconducting state of the tricolor superlattices in parallel field ($\bm{H}\parallel ab$).  Figure\,6(a) shows the $T$-dependence of $H_{c2\parallel}/T_c$ for the tricolor superlattices, along with the data for  CeCoIn$_5$ single crystal and the bicolor superlattices with the same CeCoIn$_5$ BL thickness.   It is obvious that $H_{c2\parallel}/T_c$ of  the tricolor superlattices is largely enhanced from those of the bicolor superlattices  and  single crystal.   Similar to the perpendicular field case, this enhancement can be attributed to the Rashba effect, because in-plane field component which satisfies  $\bm{H}\perp\bm{g}(\bm{k})$ for certain momenta $\bm k$ does not cause  the Zeeman splitting, giving rise to the reduction of the Pauli paramagnetic effect.

\begin{figure*}[t]
	\begin{center}
		\includegraphics[width=0.7\linewidth]{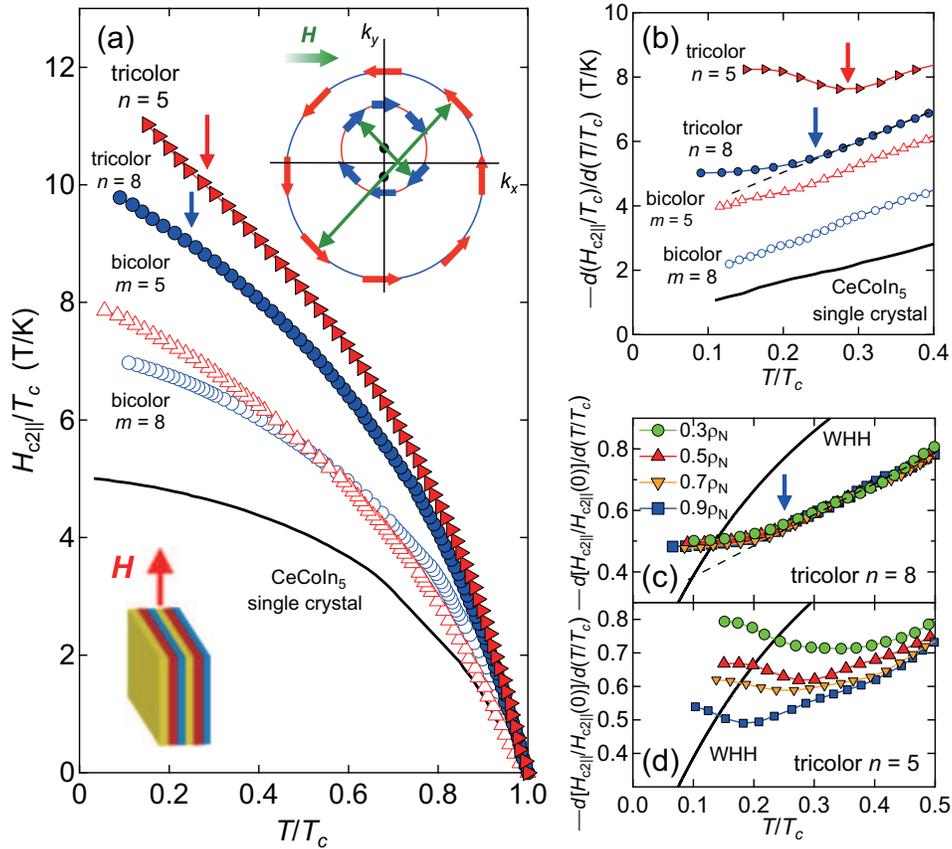}
		\caption{
		(a) Reduced upper critical field in parallel fields, $H_{c2\parallel}/T_c$, as a function of $T/T_c$ for the tricolor superlattices, compared with that for the bicolor CeCoIn$_5$($m$)/YbCoIn$_5$(5) superlattices with the same CeCoIn$_5$ block layer thickness. 
		$H_{c2\parallel}$ is determined by using $\rho(T, H)=0.5\rho_N(T)$. 
		We also plot $H_{c2\parallel}/T_c$ for CeCoIn$_5$ single crystal. 
		Inset illustrates schematic figure of the Fermi surfaces under parallel magnetic fields. Arrows on the Fermi surfaces indicate spins. Fields parallel to $x$ axis shifts the center of the Rashba-split small and large Fermi surfaces by $\bm{q_M}$ along $+y$ and $-y$ directions, respectively. Pairing occurs between the states of $\bm{k}+\bm{q_M}$ and $-\bm{k}+\bm{q_M}$, leading to a gap function with spatial modulation, e.g., in a form of $\Delta(\bm{r})=\Delta_0e^{i\bm{q_M}\cdot\bm{r}}$ \cite{Agterberg}.
		(b) Temperature derivative of $H_{c2\parallel}$, represented as $r=-d(H_{c2\parallel}/T_c)/d(T/T_c)$, is plotted as a function of $t=T/T_c$. $r(t)$ shows a minimum for the $n=5$ tricolor superlattice (red arrow). For $n=8$, $r(t)$ shows a deviation from $T$-linear temperature dependence (dashed line) below $t\sim 0.25$ (blue arrow). 
		(c) and (d) shows $r(t)$, replotted as $-d[H_{c2\parallel}/H_{c2\parallel}(0)]/d(T/T_c)$, for (c) $n=5$ and $n=8$ tricolor superlattices, respectively. Here we use $H_{c2\parallel}$ determined by using different criteria, $\rho(T,H)=0.3\rho_N(T)$, $0.5\rho_N(T)$, $0.7\rho_N(T)$ and $0.9\rho_N(T)$. Solid lines are $r(t)$ calculated from the WHH formula.
		}
	\end{center}
\end{figure*}

Figure\,6(a) shows that each system exhibits characteristic $T$-dependence of $H_{c2\parallel}$.   This can be seen clearly in Fig.\,6(b), which plots $r=-d(H_{c2\parallel}/T_c)/d(T/T_c)$ against $t=T/T_c$.   In CeCoIn$_5$ single crystal, $r(t)$  decreases linearly with decreasing $t$ and goes to zero  at $t \rightarrow 0$, indicating that  $H_{c2\parallel}$ tends to saturate at low temperatures.    In $m=5$ and 8 bicolor superlattices,  $r(t)$  decreases linearly with finite residual value at $t\rightarrow 0$. Markedly different temperature dependence of $H_{c2\parallel}$ is observed in the tricolor superlattices. In $n=8$ superlattice, $r(t)$ shows a deviation from $T$-linear behavior below $t\approx0.25$. For $n=5$, $r(t)$ increases after showing a distinct minimum  at $t\approx0.28$ as $t$ is lowered. These results indicate the upturn behavior of $H_{c2\parallel}(T)$ at low temperatures in the tricolor superlattices [see red and blue arrows in Figs.\,6(a) and 6(b)]. We stress that the observed upturn behavior of $H_{c2\parallel}(T)$ is intrinsic because of the following reasons. First of all, Figs. 6(c) and 6(d) show $r(t)$ determined by four different criteria [$\rho(T,H)=0.3$, 0.5, 0.7, and $0.9\rho_N(T)$] for $n=8$ and 5 tricolor superlattices, respectively. For $n=8$, all $r(t)$ data collapse into a single curve, indicating that the deviation from $T$-linear behavior is independent of the criteria used. Although the data for $n=5$ do not collapse into a single curve, the fact that all $r(t)$ curves exhibit increasing behavior at low temperature suggests that the upturn of $H_{c2\parallel}(T)$ occurs at the onset, middle, and tail parts of the resistive transition. Secondly, as shown in Fig.~4(d), the resistive transition taken from 4\,T to 5.5\,T with an interval of $\Delta H=0.1$\,T  around $\rho(T,H)=0.5\rho_N(T)$ exhibits a parallel shift with similar change in $T_c$, $\Delta T_c$, indicating that $\Delta T_c/\Delta H$ is constant even at low $T$. This is in sharp contrast to the other superconductors in which $\Delta T_c/\Delta H$ increases rapidly as $T\rightarrow0$. Thirdly, in Figs. 6(c) and 6(d), $r(t)$ of tricolor superlattices is compared with that of the WHH curve with no Pauli pair-breaking effect. We note that $r(t)$ of conventional superconductors cannot exceed $r(t)$ of the WHH curve. Remarkably, in both $n = 8$ and 5 tricolor superlattices, $r(t)$ exceeds that of the WHH curve irrespective of the $H_{c2}$ criteria. Based on these results, we conclude that the upturn behavior of $H_{c2\parallel}(T)$ is a unique property of the tricolor superlattices.

We note that the upturn behavior of $H_{c2\parallel}(T)$ is not caused by the multiband or strong coupling effect, because both of them give positive curvature of upper critical field immediately below $T_c$ \cite{Gurevich,Bulaevskii}.   We point out that the upturn of upper critical field at low temperature may be a signature of new superconducting phase.  It is well known that  such an upturn occurs  by a formation of the Fulde-Ferrell-Larkin-Ovchinnikov (FFLO) state \cite{Matsuda2002}, in which the pairing occurs between the Zeeman-split part of the Fermi surface, as  reported in layered organic superconductors in parallel field \cite{Lortz}.  The FFLO state  is characterized by the formation of Cooper pairs with nonzero total momentum $({\bm k}+{\bm q}\uparrow,\,  -{\bm k}+{\bm q}\downarrow)$ instead of the ordinary BCS pairs $({\bm k}\uparrow,\,-{\bm k}\downarrow)$.  In the lowest Landau level solution, superconducting order parameter is spatially modulated as $\cos({\bm q}\cdot{\bm r})$ with  ${\bm q}\parallel{\bm H}$ \cite{Matsuda2002}.   However, the FFLO state is highly unlikely to be the origin of upturn because  the Rashba splitting well exceeds the Zeeman energy ($\alpha_R\gg\mu_B H $) in the present tricolor superlattices. 

Recently  the helical and stripe superconducting states have been proposed in 2D superconductors with global inversion symmetry breaking in magnetic field parallel to the 2D plane \cite{Kaur,Agterberg}. These states appear as a result of the shift of the Fermi surface with the Rashba SOI by the external magnetic field.  When the magnetic field is applied along the $\hat{{\bm x}}$ axis ({\boldmath $H$}=$H\hat{{\bm x}}$), the centers of the two Fermi surfaces with different spin helicity are shifted along  $\hat{{\bm y}}$ in opposite directions, as illustrated in the inset of Fig.\,6(a).  Similar to the FFLO state, these states are characterized by the formation of Cooper pairs $({\bm k}+{\bm q_M},\, -{\bm k}+{\bm q_M})$, where ${\bm q_M}=m_q\mu_BH\hat{{\bm y}}/|\bm{k}|$ $(\perp{\bm H})$ with $m_q$ the mass of the quasiparticles.  The phase of superconducting order parameter is  modulated as $\Delta({\bm r})=\Delta_0 e^{i\bm{q_M}\cdot{\bm r}}$ in the helical superconducting state and $\Delta({\bm r})=\Delta_1 e^{i\bm{q_M}\cdot{\bm r}}+\Delta_2 e^{-i\bm{q_M}\cdot{\bm r}}$ in the stripe state.  It has been shown that the formation of helical and stripe superconducting states enhances $H_{c2\parallel}$ at low temperature, giving rise to the upturn behavior of $H_{c2\parallel}$ \cite{Kaur,Agterberg}.  
		
Although the presence of helical or stripe phases should be scrutinized, the facts that anomalous upturn of $H_{c2\parallel}$ is observable only in the tricolor superlattices and  is more pronounced in the superlattices with smaller $n$  imply that the Rashba SOI induced by the global inversion symmetry breaking plays an essential role in producing a high field superconducting phase.   More direct measurements which sensitively detect the change of superconducting order parameter, such as scanning tunneling microscope \cite{Allan,Zhou} and site-selective nuclear magnetic resonance \cite{Yamanaka,Kakuyanagi,Kumagai}, are strongly desired.

\section{Summary}
 
We have designed and fabricated tricolor Kondo superlattice, in which strongly correlated 2D superconductor CeCoIn$_5$ is sandwiched by nonmagnetic metals YbCoIn$_5$ and YbRhIn$_5$ with different electronic structure.   By stacking three compounds repeatedly in an asymmetric sequence such as (YbCoIn$_5$/CeCoIn$_5$/YbRhIn$_5$)$_{\ell}$,  we can introduce the global inversion symmetry breaking  along the stacking direction. We find that the Rashba SOI induced by the global inversion symmetry breaking in these tricolor Kondo superlattices leads to profound changes in the superconducting properties of CeCoIn$_5$ with atomic thickness. The upper critical field exhibits unusual temperature and angular dependence, which are essentially different from those in CeCoIn$_5$ single crystals. These results indicate that the Rashba effect induced in the tricolor superlattices leads to the strong suppression of the Pauli paramagnetic pair-breaking effect. We also demonstrate that the magnitude of the Rashba SOI incorporated into the 2D CeCoIn$_5$ BLs is largely controllable by changing the thickness of CeCoIn$_5$ BLs. 

Bulk CeCoIn$_5$ has 3D anisotropic electronic structure and hosts  an abundance of fascinating superconducting properties.   The $d_{x^2-y^2}$ symmetry is well established \cite{Izawa,Park,Stock,An,Allan,Zhou}.   At low temperature in magnetic field applied parallel to the $ab$ plane, a possible appearance of exotic superconducting phase, such as FFLO \cite{Radovan,Bianchi,Kakuyanagi,Matsuda2002,Kumagai}, $\pi$-triplet pairing state \cite{Yanase2009}, and coexistence of superconductivity and field-induced magnetically ordered phase ($Q$-phase) \cite{Young,Kenzelmann} has attracted intense attention in recent years in an effort to search for exotic pairing states.    It has been recently pointed out that in 2D noncentrosymmetric superconductors with nodes in the superconducting gap, the topological superconducting state is stabilized with no fine tuning of parameters, in contrast to those without nodes \cite{Daido,Yuan}. Therefore, the fabrication of tricolor superlattices containing $d$-wave superconducting layers offers the prospect of achieving even more fascinating pairing states than bulk CeCoIn$_5$, such as helical and stripe superconducting states \cite{Agterberg}, a pair-density-wave state \cite{Yoshida2012}, complex stripe state \cite{Yoshida2013}, a topological crystalline superconductivity \cite{Yoshida2015}, and Majorana fermion excitations \cite{Yuan}, in strongly correlated electron systems.

\section*{Acknowledgements}
We thank Y. Yanase for valuable discussions. This work was supported by Grants-in-Aid for Scientific Research (KAKENHI) (No.~25220710, 15H02014, 15H02106, 15H05457), and Grants-in-Aid for Scientific Research on innovative areas ``Topological Materials Science" (No.~15H05852) and ``3D Active-Site Science" (No.~26105004) from Japan Society for the Promotion of Science (JSPS).

\end{document}